\documentclass[sigconf]{acmart}
\usepackage{algorithmic}
\usepackage{graphicx}
\usepackage{textcomp}
\usepackage{xcolor}
\usepackage{booktabs}
\usepackage{makecell}
\usepackage{subcaption}
\usepackage{multirow}
\usepackage{tcolorbox}
\usepackage{adjustbox}
\usepackage{enumitem}
\AtBeginDocument{%
  \providecommand\BibTeX{{%
    \normalfont B\kern-0.5em{\scshape i\kern-0.25em b}\kern-0.8em\TeX}}}

\copyrightyear{2024}
\acmYear{2024}
\acmDOI{XXXXXXX.XXXXXXX}
\acmConference[]{}{}{}
\acmPrice{15.00}
\acmISBN{978-1-4503-XXXX-X/18/06}

\setcopyright{none}
\renewcommand\footnotetextcopyrightpermission[1]{} %
\begin{document}

\title{Harnessing Integrated CPU-GPU System Memory for HPC: \\a first look into Grace Hopper}

\author{Gabin Schieffer}
\orcid{0009-0003-6504-7109}
\affiliation{\institution{KTH Royal Institute of Technology}\country{Sweden}}
\email{gabins@kth.se}

\author{Jacob Wahlgren}
\affiliation{\institution{KTH Royal Institute of Technology}\country{Sweden}}
\email{jacobwah@kth.se}

\author{Jie Ren}
\affiliation{\institution{William \& Mary}\country{USA}}
\email{jren03@wm.edu}

\author{Jennifer Faj}
\affiliation{\institution{KTH Royal Institute of Technology}\country{Sweden}}
\email{faj@kth.se}

\author{Ivy Peng}
\affiliation{\institution{KTH Royal Institute of Technology}\country{Sweden}}
\email{ivybopeng@kth.se}

\begin{abstract}
Memory management across discrete CPU and GPU physical memory is traditionally achieved through explicit GPU allocations and data copy or unified virtual memory. The Grace Hopper Superchip, for the first time, supports an integrated CPU-GPU system page table, hardware-level addressing of system allocated memory, and cache-coherent NVLink-C2C interconnect, bringing an alternative solution for enabling a Unified Memory system. In this work, we provide the first in-depth study of the system memory management on the Grace Hopper Superchip, in both in-memory and memory oversubscription scenarios. We provide a suite of six representative applications, including the Qiskit quantum computing simulator, using system memory and managed memory. Using our memory utilization profiler and hardware counters, we quantify and characterize the impact of the integrated CPU-GPU system page table on GPU applications. Our study focuses on first-touch policy, page table entry initialization, page sizes, and page migration. We identify practical optimization strategies for different access patterns. Our results show that as a new solution for unified memory, the system-allocated memory can benefit most use cases with minimal porting efforts.

\end{abstract}


\maketitle

\section{Introduction}
Graphics Processing Units (GPUs) have become pivotal in parallel computing and high-performance computing (HPC). Today, most top supercomputers on TOP500 are accelerated with GPUs. The massive parallelism, high computing throughput, and power efficiency brought by GPUs, make them crucial for compute-intensive tasks in parallel scientific applications, image processing, and machine learning, enabling advancements in AI research, data science, and scientific and engineering challenges.

Memory management across discrete CPU and GPU physical memory is traditionally achieved through explicit GPU allocations and data copy. Moreover, workloads on GPUs, such as large language models (LLM), quantum computer simulators, and particle simulations, often require significant memory capacity to store and process massive datasets. However, GPUs, despite being essential for accelerating computationally intensive tasks, have limited high-bandwidth memory in the order of tens of gigabytes. In contrast, today's CPU memory on high-end platforms is often in the order of hundreds of gigabytes. To address the memory capacity bottleneck faced by GPU-accelerated applications, existing solutions such as Unified Virtual Memory (UVM)~\cite{unifiedmem} and data object offloading~\cite{atc21:zerooffload} are proposed to provide software-level approaches that extend GPU memory capacity by utilizing CPU memory.

Despite their benefits, existing solutions have limitations that can impact performance and usability. For instance, UVM incurs large overheads in handling page faults in GPU and suffers from read/write amplification due to page-level swapping. Data object offloading requires offline profiling and application refactoring, limiting solution generality. Moreover, the performance of both solutions is constrained by data transfer bottlenecks between the CPU and GPU, as communication latency and bandwidth limitations hinder the overall execution speed~\cite{lutz2020pump}.

The introduction of the Nvidia Grace Hopper Superchip presents a new opportunity to address the limitations of existing solutions. The system interconnects an ARM CPU with an Nvidia H100 GPU by a cache-coherent interconnect, NVLink-C2C (chip-to-chip). In this system, a single virtual memory space is shared between the CPU and GPU (i.e., \textit{system memory}), and address translation is accelerated by hardware. This enables application developers to transparently manage memory across CPU and GPU, while delegating data transfers to hardware into two levels, i.e., direct remote accesses at cacheline granularity, and heuristic-guided page migrations. By leveraging cacheline level access and Address Translation Service (ATS), which enables full access to all CPU and GPU memory allocations, the system memory eliminates the page-fault handling overhead needed in \textit{managed memory} in UVM, and minimizes the need for memory migrations. 
While managed memory splits the virtual memory space into both the system page table and GPU page table, system memory relies on a single system-wide page table, shared between the CPU and the GPU.

Currently, there is a lack of comprehensive studies on memory allocation, memory management, and page migration overhead in the new CPU-GPU coherent system memory on Grace Hopper and the impact of the integrated system page table. Understanding these aspects is crucial for developers and researchers to fully harness the potential of the first hardware-accelerated Unified Memory system on a CPU-GPU platform. This paper aims to bridge this gap and provide insights into the performance implications of the new Unified Memory solution -- system memory, as compared to the existing alternative, managed memory, in six representative HPC applications, including the state-of-the-art Qiskit quantum computer simulator, graph, and scientific applications. 

The contributions of this paper are summarized as follows:
\begin{itemize}[leftmargin=*,topsep=5pt]
    \item We provide an in-depth analysis of the CPU-GPU integrated page table on the Grace Hopper Superchip and its implication
    \item We provide a set of six HPC applications including the Qiskit quantum computer simulator, using the new system allocated memory and CUDA managed memory
    \item We study the memory utilization behavior using a memory profiler and hardware counters on Grace Hopper in both in-memory and memory oversubscription
    \item We characterize the impact of first-touch policy, page table entry initialization, system page size, and page migration mechanisms in system memory and managed memory on Grace Hopper
\end{itemize}

\begin{figure*}
    \centering
    \includegraphics[width=0.75\linewidth]{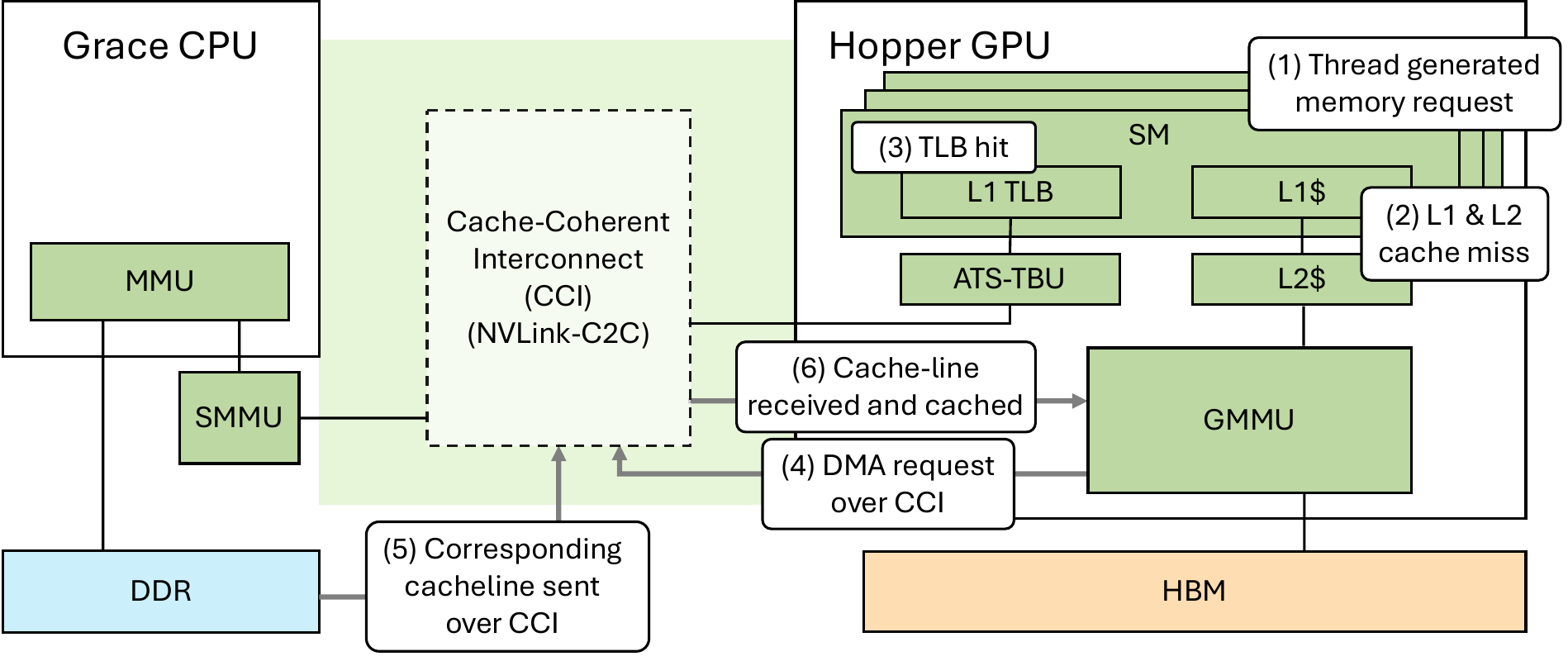}
    \caption{An overview architecture of the Grace Hopper platform that interconnects CPU and GPU with high-throughput cache-coherent NVLink-C2C.}
    \label{fig:arch}
\end{figure*}

\section{Grace Hopper Unified Memory System}
\label{sec:bg}

In this section, we introduce the memory subsystem of the Grace Hopper Superchip, focusing on hardware- and Operating System- (OS) level memory management within the single shared CPU-GPU memory domain. We further discuss how this memory system is exposed to programmers through two types of unified memory management solutions: system-allocated memory and CUDA managed memory.

\subsection{Memory Subsystem}
\label{sec:bg-1}
The Grace Hopper system features a two-tier memory architecture, with distinct physical memory regions respectively attached to the Grace CPU and the Hopper GPU. The CPU is connected to 480 GB of LPDDR5X memory, while the GPU is equipped with 96 GB of HBM3 memory.
These two processors, GPU and CPU, are interconnected via the Nvidia NVLink-C2C interconnect. The two-tier memory system is exposed as two non-uniform memory access (NUMA) nodes, allowing seamless access to both CPU and GPU memory.

We evaluated the performance of this memory architecture with benchmark tests. Using the STREAM benchmark, we measured memory bandwidth. Results show that the GPU's HBM3 memory achieved a bandwidth of 3.4 TB/s, compared to its theoretical bandwidth of 4 TB/s. The CPU's LPDDR5X memory reached a bandwidth of 486 GB/s, close to its theoretical bandwidth of 500 GB/s~\cite{gh_whitepaper}.
Furthermore, using the Comm|Scope benchmark~\cite{pearson2019evaluating}, we assessed the performance of the NVLink-C2C interconnect. We achieved a bandwidth of 375 GB/s for host-to-device (H2D) transfers and 297 GB/s for device-to-host (D2H) transfers, compared to the interconnect's theoretical bandwidth of 450 GB/s.

\subsubsection{NVLink-C2C Interconnect}
\label{sec:bg-c2c}

In the Grace Hopper system, a processor (CPU or GPU) can directly access the other processor's physical memory over the NVLink-C2C interconnect. This access is performed at the cacheline granularity, with transfer sizes as small as 64 bytes on the CPU side and 128 bytes on the GPU side.
The accessed memory is transparently cached in the cache hierarchy of both processors, as shown in Figure~\ref{fig:arch}. Additionally, the two processors' caches are fully coherent. Atomic operations can also be issued on the NVLink-C2C interconnect, allowing any processor to atomically read and modify a physical memory location. These features are implemented at the hardware level and do not require user intervention, following Arm's AMBA CHI protocol.

\subsubsection{System-level Address Translation}

In the Grace Hopper memory system, the Grace CPU features a unique hardware unit called the System Memory Management Unit (SMMU)~\cite{gh_tuning}, defined in Arm's SMMUv3 specification. The SMMU is responsible for translating virtual addresses to physical addresses by performing page table walks. 
Compared to a traditional MMU, the SMMU provides additional support for virtual-to-physical address translation requests from the GPU.

Figure~\ref{fig:arch} illustrates the workflow of an access to a virtual address location on the Grace Hopper system, when the virtual-to-physical mapping is cached in the GPU TLBs, in the situation where the data is CPU-resident.
(1) A GPU thread accesses a virtual address.
(2) The data is not cached in the GPU cache hierarchy. This generates a cache miss.
(3) The virtual address is looked up in the GPU TLBs (Translation Lookaside Buffers) for virtual-to-physical translation. As the translation is already cached, it is used to perform an access to physical memory.
(4) The GMMU initiates a direct memory access (DMA) over the NVLink-C2C interconnect, at the cacheline granularity.
(5) The requested access is performed from CPU memory, and send back to the GPU.
(6) The access is completed, and memory is cached in the regular GPU cache hierarchy.

Compared to pre-Grace Hopper systems, which rely on GPU page fault handling to access CPU memory, this new approach has two main implications. First, GPU accesses to CPU-located memory no longer systematically trigger GPU page faults. Second, page faults are now generated by the SMMU and can be directly handled by the operating system's page fault handling mechanism, simplifying the overall process.

\subsubsection{Memory Management in Grace Hopper}
\label{sec:bg-2}
\begin{table}[bt]
\caption{A list of Memory Management Types}
\label{tab:api}
\centering
\begin{adjustbox}{width=\linewidth,center}
    \begin{tabular}{cccccc}
    \toprule
    \makecell{Memory \\Location} & \makecell{Allocation \\Interface} &\makecell{PTE \\Init} & \makecell{Cache \\Coherent} & \makecell{Migration \\Granularity}\\
    \midrule
    \midrule
    CPU/GPU & \texttt{malloc()} &CPU &Yes & \makecell{transparent\\ 128 byte \\ 64KB}\\\midrule
    CPU/GPU & \makecell{ \texttt{cudaMallocManaged()} } &CPU & Yes &\makecell{transparent\\2MB} \\\midrule
    GPU & \makecell{\texttt{cudaMalloc()} \\ \texttt{cuMemCreate()}} &GPU & No &\makecell{explicit\\1 byte} \\\midrule
    CPU & \makecell{ \texttt{numa\_alloc\_onnode()} \\ \texttt{cudaMallocHost()} \\ \texttt{cudaHostAlloc()} \\ \texttt{cuMemCreate()}} &CPU & No &\makecell{explicit\\1 byte} \\\bottomrule
    \end{tabular}
\end{adjustbox}
\end{table}

The Grace Hopper system utilizes two distinct page tables: a system-wide page table and a GPU-exclusive page table. The choice of page table for a memory allocation depends on the specific programmer-level API used. Table~\ref{tab:api} summarizes memory allocation APIs on a Grace Hopper system into three categories: allocations in CPU physical memory only, allocations in GPU physical memory only, and allocations that can reside in either CPU or GPU physical memory.

\textbf{An Integrated System-wide Page Table}. 
The Grace Hopper system introduces a system-wide page table, located in CPU memory. The operating system directly accesses this page table, creates and manages page table entries (PTEs). The SMMU uses this page table to provide virtual-to-physical address translation for both the CPU (when required by user applications) and the GPU (when requested over the NVLink-C2C interconnect). Memory pages in the system-wide page table can be physically located in either CPU or GPU memory, and they use the system page size, which is defined at the operating system level and constrained by the CPU architecture capabilities. When using the Grace CPU, the page size is either 4~KB or 64~KB.

\textbf{A GPU-exclusive Page Table}. 
The Grace Hopper system retains the local GPU page table from previous generations of Nvidia GPUs. This page table, located in GPU memory and only accessible by the GPU, stores virtual-to-physical translations for \texttt{cudaMalloc} allocations and \texttt{cudaMallocManaged} allocations when the physical location of the managed memory is on the GPU. The page size used by this page table is 2~MB.

In this work, we focus on allocations where data can be resident in either CPU or GPU physical memory, namely \textit{system-allocated memory} and \textit{CUDA managed memory}, as detailed in the following sections.

\subsection{System-Allocated Memory}
System-allocated memory refers to memory allocated using standard methods like the C standard library function \texttt{malloc()}, which exclusively uses the system page table and relies on the operating system to manage the virtual memory space of a process.

In general, when \texttt{malloc} is called, the operating system creates page table entries in the system page table without assigning physical memory to those pages. This allows for over-subscription and improves performance by mapping only accessed pages to physical memory. During the first access to a virtual address in the allocation, known as \textit{first-touch}, a page fault is triggered since the accessed virtual page is not mapped to physical memory. The operating system handles this page fault by identifying unused physical memory for the requested page, updating the page table, and replaying the memory access. On Grace Hopper, this process applies to both CPU and GPU first-touch accesses.

When a GPU thread generates a first-touch access to a virtual address, a GPU TLB miss is triggered. As a result, the GPU's ATS-TBU (Translation Buffer Unit) generates an address translation request and sends it to the SMMU over NVLink-C2C. To answer the request, the SMMU performs a page table walk in the system page table. If no physical memory is allocated to the page, the SMMU issues a page fault. OS handles the fault by updating the page table entry to point to GPU physical memory, as the first-touch originated from a GPU thread. Once the physical address is stored in the GPU's TLB, GPU threads can perform memory access using direct memory access to the physical memory address, potentially located in CPU memory, over NVLink-C2C, as described in Section~\ref{sec:bg-c2c}, and pictured in Figure~\ref{fig:arch}.

\subsubsection{Automatic Delayed Access-counter-based Migrations} 
\label{sec:bg-access-counters}

In order to improve performance of applications which use system-allocated memory, the Grace Hopper system can be configured to automatically migrate memory regions between GPU and CPU physical memory~\cite{gh_whitepaper}.%
It is important to note here that this migration is different and independent from AutoNUMA migrations used in the Linux kernel.
Thanks to the fully-coherent GPU-CPU memory and direct memory access capabilities of Grace Hopper, this automatic migration feature is purely aimed at enhancing performance and does not intend to overcome technical limitations.

The default migration strategy, detailed in Nvidia's open-source GPU driver\footnote{\url{https://github.com/NVIDIA/open-gpu-kernel-modules/}}, relies on hardware counters to track GPU accesses to memory ranges.%
When a counter value exceeds a user-defined threshold (by default, 256), the GPU issues a \textit{notification} in the form of a hardware interrupt, which is handled by the GPU driver on the CPU. The driver then determines whether to migrate the pages belonging to the associated virtual memory region.
The intent of this strategy is to automatically migrate pages that are heavily accessed by the GPU from CPU memory to GPU memory, thereby improving performance.

\subsection{CUDA Managed Memory}

CUDA managed memory, introduced since CUDA 6.0, aims to provide a single virtual address space shared between CPU and GPU while still relying on two distinct page tables, one for the CPU and another for the GPU. CUDA managed memory is primarily a \textit{software} abstraction, implemented as part of the CUDA runtime libraries and the Nvidia GPU driver.

Programmers create managed memory allocations using the \texttt{cudaMallocManaged()} function. Similar to \texttt{malloc}, for post-Pascal systems, the virtual memory is not immediately mapped to physical memory. Instead, the location of the first-touch triggers this mapping operation. The performance of CUDA managed memory highly depends on hardware of the underlying system.

\subsubsection{On-demand page migration}
CUDA managed memory relies on \textit{on-demand} page migration to enable both GPU and CPU to access the shared virtual memory range.
When the GPU tries to access a page, a page fault is triggered if a GPU TLB miss occurs and the GMMU fails to find the virtual address in the GPU-exclusive page table.
This page fault causes a page migration from CPU memory to GPU memory. Meanwhile, when GPU memory is overwhelmed, pages can also be evicted to CPU memory.
A similar page retrieval process occurs for CPU access to GPU-resident memory.

Coherent dynamic memory allocation was introduced on Power9 platforms~\cite{power9} in CUDA 9.2. This feature is supported by the ATS, which enables hardware-level address translations by allowing direct communication between CPU and GPU MMUs and eliminates the need for software-level address translation.

\subsubsection{Speculative prefetching}
In addition to on-demand page migration, speculative prefetching strategies are used to migrate pages before they are accessed, in order to reduce the page fault handling overhead of CUDA managed memory on the critical path. These strategies include explicit prefetching, triggered through the \texttt{cudaMemPrefetchAsync} API, and implicit prefetching performed by GPU hardware prefetchers~\cite{ganguly2019interplay}. 
Although these prefetching mechanisms are not technically required on the Grace Hopper system, as the direct memory access capabilities allow for low-overhead access to remote memory at a cacheline granularity without triggering page faults, our experiments showed that prefetching mechanisms are still employed when using CUDA managed memory.

\section{Methodology}
\label{sec:method}
We use a Grace Hopper system as our testbed. It consists of a Grace CPU, i.e., a 72-core ARM Neoverse V2 CPU, and an Nvidia H100 GPU. The CPU has 480~GB LPDDR5X memory, and the GPU has 96~GB HBM3 memory. This system executes RHEL 9.3  with CUDA 12.4, with Nvidia GPU driver 550.54.15. The system is configured by~\cite{gh_tuning}: (1) Automatic NUMA Scheduling and Balancing is disabled because the additional page-faults introduced by AutoNUMA can significantly hurt GPU-heavy application performance; (2) New allocations are not newly allocated pages and heap objects with zeroes by default and CONFIG\_INIT\_ON\_ALLOC\_DEFAULT\_ON is off and init\_on\_alloc=0 parameter. Furthermore, we set the page migration notification threshold to the default value of 256.%

\begin{table}
\caption{A summary of applications, access patterns, and input problems ($x \times y$ indicates a 2D input problem).}
\label{tab:workloads}
\begin{adjustbox}{width=\linewidth,center}{
\centering
\begin{tabular}{|l|l|c|c|c|c|c|c}
 \hline
 \textbf{Name} &\textbf{Description}  &\textbf{Pattern} &\textbf{Input}\\\hline\hline

Qiskit~\cite{QiskitCommunity2017} & Quantum Volume Simulation & Mixed & 30-34 qubits\\\hline
Needle         & Needleman-Wunsch algorithm & Irregular   & 32k $\times$ 32k\\\hline
Pathfinder & 2-D grid pathfinding algorithm & Regular   & 100k $\times$ 20k\\\hline
BFS        & Graph processing problem. Breadth-first search & Mixed     & 16M nodes \\\hline
Hotspot    & Differential equation solver for thermal simulation & Regular   & 16k $\times$ 16k\\\hline
SRAD       & Speckle Reducing Anisotropic Diffusion & Irregular & 20k $\times$ 20k\\\hline

\end{tabular}
}\end{adjustbox}
\end{table}
\subsection{Applications}
We select a set of six applications representing different access patterns in HPC applications. This selection includes five applications from the Rodinia benchmark suite~\cite{che2009rodinia}: bfs, needle, pathfinder, hotspot, and srad. We further include in our evaluation the state-of-the-art Qiskit quantum computer simulator~\cite{QiskitCommunity2017}. Table~\ref{tab:workloads} presents those applications, along with access patterns found in the literature, and problem sizes. Three different access patterns are described in this table: \textit{regular}, which refers to applications with dense memory accesses to contiguous virtual address ranges; \textit{irregular}, which refers to applications performing sparse accesses over a large range virtual addresses; and mixed, for applications exhibiting both irregular and regular behaviors over different regions. Those access patterns are presented in details in~\cite{go2023early}. The Quantum Volume simulation uses a state-vector quantum simulator, which exhibits a mixed access pattern. This simulator is evaluated on the previous generation of Nvidia GPUs, without NVLink-C2C interconnect, in~\cite{qiskita1002023}. In this setup, the CPU-GPU data transfers are identified to be a major performance bottleneck in large-qubits scenarios, which makes the Grace Hopper Superchip an entailing platform to execute such workload.

We derive two versions for each application, one using CUDA managed memory and one using system-allocated memory. For this purpose, we first identify candidate memory allocations to replace, by locating explicit host-to-device data movements in the code. 
We replace the destination and source buffers in those data transfers by a single buffer, allocated using one of the two unified memory allocators, either the system-level allocator (\texttt{malloc}) or CUDA managed memory allocator (\texttt{cudaMallocManaged}). GPU-only buffers, which are never meant be accessed by the CPU, and are typically only used for storing intermediary results on the GPU, are still allocated with \texttt{cudaMalloc}.

Removing explicit data transfers also effectively removes some synchronization points, we instead add explicit CUDA device synchronization calls, in order to ensure that the semantics of the application is preserved, and that no race condition is present.

\begin{figure}
    \centering
    \includegraphics[width=0.9\linewidth]{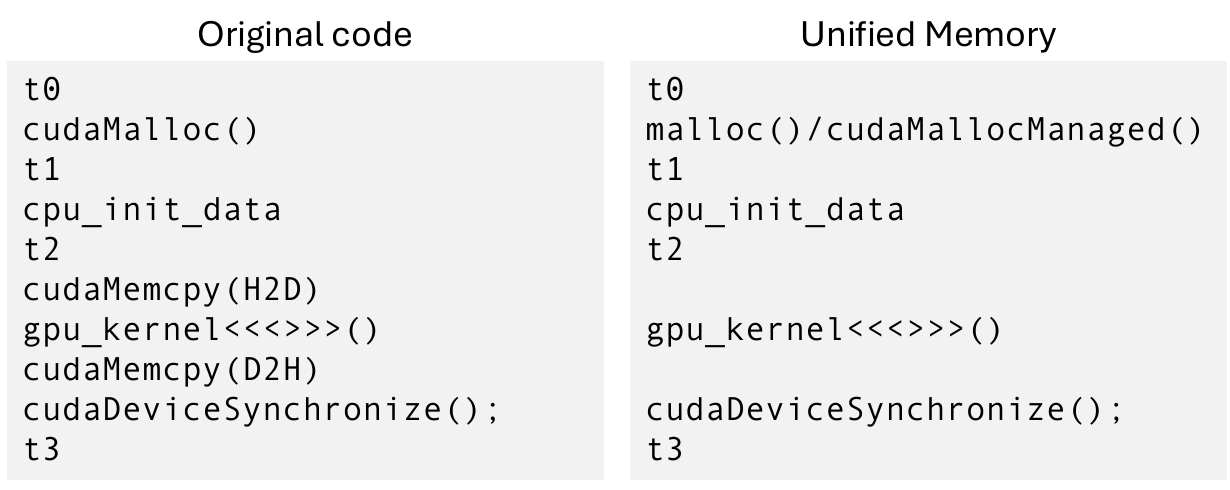}
    \caption{A snippet of code transformation from a typical CUDA code with explicit memory copy to Unified Memory.}
    \label{fig:timing}
\end{figure}

For small-scale benchmarks, we use CPU timers to measure execution of the various execution phases of each application, through the C standard library function \texttt{gettimeofday}. This timing method has a low overhead ($<1~\mu s$), compared to the measured quantities ($>1~ms$).
To provide consistent measurements across different versions of the same application, and across different applications, we established several phases which are common to all application: GPU context initialization and argument parsing, allocation, CPU-side buffers initialization, computation, and de-allocation. Figure~\ref{fig:timing} presents a comparison in pseudo-code of original code and our modified version, on this figure, the duration for allocation, CPU-side initialization, and computation phases are respectively $t_1-t_0$, $t_2-t_1$, $t_3-t_2$.

In Rodinia benchmarks, the CPU-side initialization is single-threaded and generally involves intensive I/O operations. In some cases, this phase represents more than 95\% of the overall application runtime. Moreover, since CPU-side initialization is only limited by CPU and I/O performance, the execution time difference for this phase is negligible for all versions of the same application. For those reasons, we exclude this phase when reporting absolute timing.

We performed the quantum computing simulator experiments with Qiskit-Aer, an open-source implementation that is capable of using GPUs as a backend through CUDA and the thrust library. As our experiments are based on a statevector simulator, the main data structure that serves as a buffer of the statevector requires $8*2^{Nqubits}$ bytes of memory. For porting to Unified Memory, we use the same approach as for the Rodinia applications. In addition, for thrust-related memory allocations, we define a custom memory allocator, using either \texttt{cudaMallocManaged()} or \texttt{malloc()}.

By default, the simulator already supports heterogeneous architectures through an explicit exchange of chunks between CPU and GPU in case the circuit's memory requirement exceeds the available memory on the GPU. In order to avoid this explicit copying in our unified memory approach, the maximum available memory is changed from only considering the GPU memory to considering the whole system's memory.

As a benchmark we use the Quantum Volume Circuit simulating up to 34 qubits, where up to 33 qubits fit into GPU memory and 34 qubits exceeds GPU memory. At the core of the simulation are a series of matrix multiplications that benefit from a high memory throughput.

\subsection{Profiling Tools}
\label{sec:tool}

In order to construct a memory profile of our applications, we develop a simple memory profiling tool. The intent is to capture the memory usage of a process for both CPU physical memory and GPU physical memory. For the CPU usage of the process, we use the resident set size (RSS), as reported on a per-process basis. Resident set size represents the number of pages which are actively used by a process, that is, which are directly mapped to physical memory; this value is accessible through the \texttt{/proc/<pid>/smaps\_rollup} interface. For the GPU memory, we use the GPU used memory value provided by nvidia-smi, which includes memory footprint for cudaMalloc, cudaMallocManaged, and system-level allocations, for GPU-resident pages. This value is system-wide, and comprises a \textasciitilde600~MB driver-induced baseline value.

The sampling period is set by the user. In our experiments, we use a sampling period of 100~ms. Memory profiles obtained with our profiling tool for hotspot and Qiskit are presented respectively in Figure~\ref{fig:memutil-hotspot}, and Figure~\ref{fig:memutil-qiskit}.

We use Nvidia Nsight Systems to identify GPU page faults, and page migration for managed memory. It is important to note here that this tool is only reliable for managed memory, as page faults in system-allocated memory are not reported.
For kernel-level characterization, we use the Memory Workload Analysis tool in Nvidia Nsight Compute to quantify memory traffic associated with each GPU kernel launch; this includes traffic over NVLink-C2C, system memory, and global GPU memory.

We use two setups for oversubscription scenarios. First, for Qiskit, as the memory footprint for $N_{qubits} \geq 34$ exceeds the available GPU memory, such problem gives a \textit{natural} oversubscription scenario. Second, we use a \textit{simulated} oversubscription scenario. This setup is used for Rodinia applications and Qiskit for when $N_{qubits} \leq 33$, as the maximal memory footprint for this scenario does not exceed the total GPU memory. 
To emulate oversubscription, we create a $N$-byte \texttt{cudaMalloc} allocation. We then measure the amount of free GPU memory $M_{gpu}$. This represents the memory which the application will be able to use on the GPU. We obtain the oversubscription ratio using $R_{oversub} = M_{peak} / M_{gpu}$, where $M_{peak}$ is the peak application GPU memory usage, measured using our memory profiling tool described in Section~\ref{sec:tool}, in a non-oversubscribed case.

\section{Overview}
\begin{figure*}
    \centering
    \includegraphics[width=0.9\linewidth]{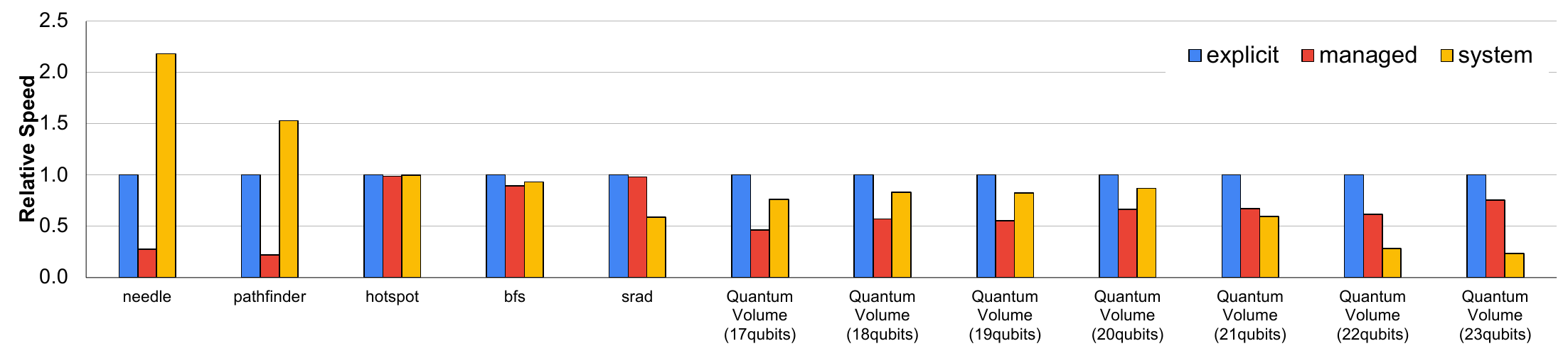}
    \caption{An overview of the relative performance of the system-allocated memory and the managed memory version, in terms of speedup, compared to the original explicit data copy version in six applications. No specific optimizations are included.}
    \label{fig:overview}
\end{figure*}
\begin{figure}[bt]
    \centering
    \includegraphics[width=\linewidth]{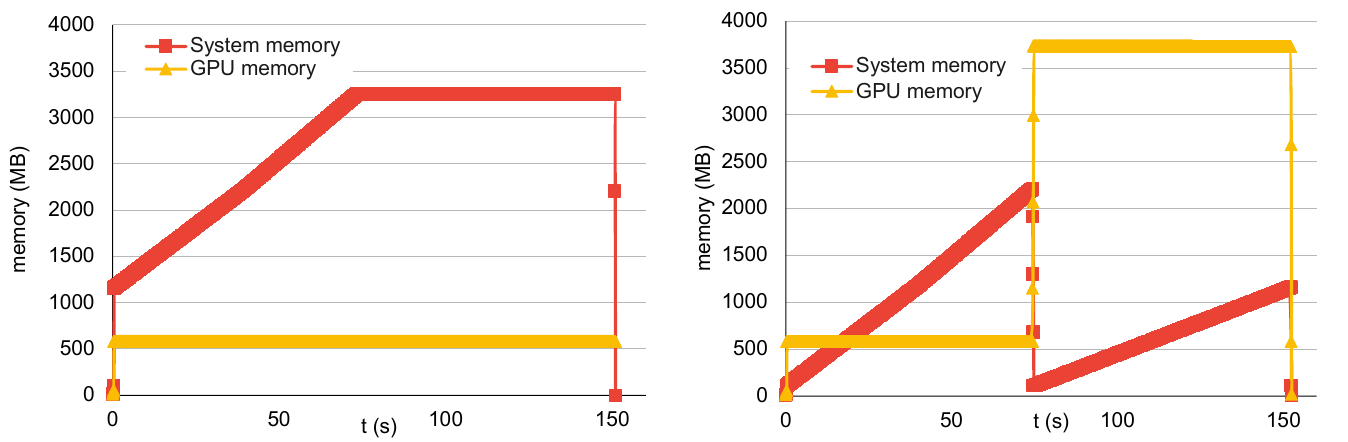}
    \caption{The memory usage patterns over time in hotspot using system memory (left) and CUDA managed memory (right).}
    \label{fig:memutil-hotspot}
\end{figure}
\begin{figure}[bt]
    \centering
    \includegraphics[width=0.9\linewidth]{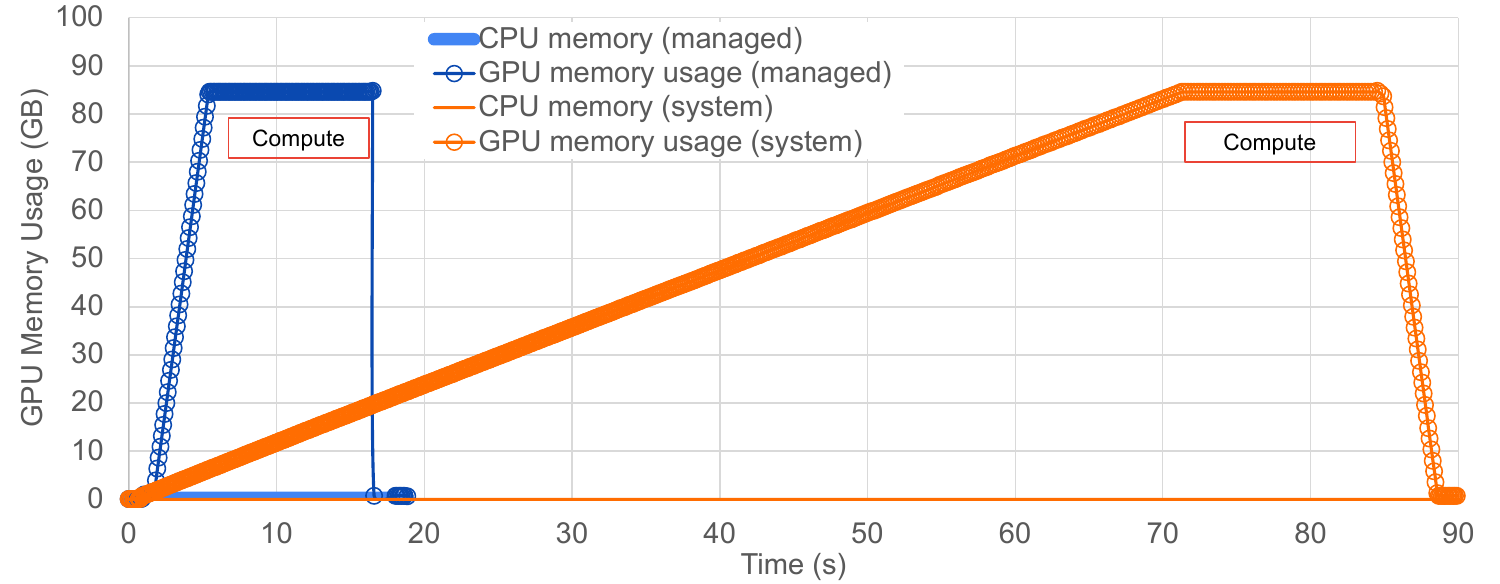}
    \caption{The memory usage patterns over time in Qiskit Quantum Volume simulation using system memory and managed memory, respectively.}
    \label{fig:memutil-qiskit}
\end{figure}

In this section, we present an overview of comparing the two Unified Memory solutions -- system-allocated memory and CUDA managed memory, with the traditional explicit copy implementation in six applications. Figure~\ref{fig:overview} presents the relative performance of each application compared to their original version that uses explicit GPU allocations and data transfer between CPU and GPU (denoted as \textit{explicit}). Note that no explicit optimizations are applied to the two versions. They represent the results after applying the code transformation described in Section~\ref{sec:method}. In these experiments, applications fit in GPU memory and automatic page migration in system memory is disabled. 

Overall, the relative performance of the system memory and managed memory versions can be categorized into two classes. In applications, such as needle, pathfinder, hotspot, bfs, and Quantum Volume simulations of 17-20 qubits, the system memory version outperforms the managed memory version. The managed memory will trigger page faults when the GPU accesses data that is not in GPU memory, and start on-demand page migration. As pointed in multiple existing works~\cite{ganguly2019interplay,allen2021demystifying}, the page fault handling can cause higher overhead than the data migration itself. The new cache-coherent NVlink-C2C enables direct data access to CPU memory at cacheline level without involving the expensive page fault mechanism, attributing to the observed speedup. For some cases, such as needle and pathfinder, the system memory version even outperforms the original explicit version. As shown in Section~\ref{sec:pagetable}, there could be a significant difference in the allocation and de-allocation time depending on the type of memory management in use.

In contrast, for SRAD and Quantum Volume simulations of 21-23 qubits, the managed memory version outperforms the system memory version. Our in-depth analysis in Section~\ref{sec:pagetable} identifies the main factors coming from the data structures that are initialized on GPU and the different sizes of the integrated system pages and GPU-exclusive pages. We note that Quantum Volume simulations have higher performance in the original version than the two unified memory versions. This is expected as the original version implements a sophisticated data movement pipeline and represents the ideal performance. As shown in Section~\ref{sec:page-migration}, with the optimization of prefetching applied into the managed memory version, we can achieve performance close to that of the explicit version. As shown in Section~\ref{sec:pagesize}, the performance of system-allocated memory improves significantly when increasing the system page size and has high dependency in initialization phase.

We also identified a difference in behavior for the GPU context initialization. In the traditional explicit version and managed memory version, memory allocations, and data transfer are done through specific CUDA APIs before kernel launches, which implicitly initialize GPU context. However, in the system memory version, due to the absence of explicit CUDA memory allocation and data copy API calls, GPU context initialization occurs within the first kernel launch, apparently prolonging the computation time.

To understand the different memory utilization patterns in the two unified memory versions, we leverage our memory profiler to characterize the six tested applications. Due to space limit, we only present two types of memory usage.

Figure~\ref{fig:memutil-hotspot} presents the memory usage over time in hotspot. In the system memory version, GPU memory usage stabilizes over the whole execution while the system memory usage slowly ramp up first till the end of initialization phase and then stabilize throughout the computation. In contrast, in the managed memory version, the system memory usage also slowly ramp up initially. However, once in the computation phase, GPU access to data triggers page migration, and a steep decrease in system memory and a sharp increase in GPU memory usage is observed. As discussed in the previous sections, hotspot represents a typical class of existing GPU applications, where data structures used in GPU computation are initialized on CPU.

Figure~\ref{fig:memutil-qiskit} presents the memory usage over time in the Quantum Volume simulation. In this application, the end-to-end execution is significantly prolonged in the system memory version, compared to the managed memory version. However, we also notice that the main difference is only constrained in the initialization phase, where the GPU memory usage slowly ramps up in the system memory version (orange) but quickly reaches the peak in the managed memory version (blue). In fact, the computation phase in both versions are similar. We present a detailed analysis in Sections~\ref{sec:gpu-first-touch} and~\ref{sec:page-migration}.

\section{CPU-GPU Integrated System Page Table}
\label{sec:pagetable}
In this section, we focus on first-touch page placement and system page size, two main aspects that can affect the impact of the CPU-GPU integrated system page table on GPU applications.  
 
\subsection{First-touch Page Placement}
System memory uses a first-touch placement policy and pages always reside in the system page table, while managed memory also uses a first-touch placement policy but pages may reside in either the system page table or GPU page table, depending on its physical location. To compare their sensitivity to the first-touch policy, we use two sets of benchmarks, representing GPU-initialized (srad and qiskit) and CPU-initialized benchmarks (bfs, hotspot, pathfinder, and needle).

\subsubsection{CPU-side initialization}
\label{sec:cpu-first-touch}
The common programming model of GPU-accelerated HPC applications is to perform data initialization, often including pre-processing, on the CPU before offloading data onto the GPU for computation. In such a pattern, the first-touch policy will cause pages to be placed on the CPU during initialization. When the computation phase starts, in managed memory, data is migrated on demand to the GPU memory often with additional pages from speculative prefetching, which will result in both traffic on the NVLink-C2C and increased GPU memory utilization. Instead, in the system memory, data will not be migrated on access but deferred, which will result in only traffic on the NVLink-C2C link and no immediately increased GPU memory utilization. Consequently, memory usage as shown in Figure~\ref{fig:memutil-hotspot} and the measured traffic over NVLink-C2C signify the difference in the two patterns. 

\subsubsection{GPU-side initialization}
\label{sec:gpu-first-touch}
When data is initialized and first touched by GPU, as in SRAD and Qiskit, CUDA managed memory and system-allocated memory exhibit significantly different behaviors. With CUDA managed memory, as shown in Figure~\ref{fig:memutil-qiskit}), the initialization is much shorter than that in the system memory version, and no page migration is performed during the computation phase, as the first touch by GPU has directly mapped data to GPU memory. With system memory, the GPU first-touch policy triggers a replayable page fault, as the page being first-touched is neither present in the GMMU page table, nor through address translation. The CPU then handles the page fault and populates the system page table, therefore slowing down the initialization time on the GPU. Consequently, the initialization phase is significantly longer, as shown in Figure~\ref{fig:qiskit_pagesize}.

Comparing the two initialization schemes, we note that system-allocated memory performs better in cases of CPU-side initialization as the page faults are both triggered and handled on the CPU side, whereas in the GPU-side initialization, page table initialization on the CPU-side significantly slows down the execution. In the latter cases, we observed that CUDA managed memory performed better. We propose new strategies that can potentially reduce the impact of the page table entry creation when using system-allocated memory. For applications with CPU-side initialization, the \texttt{cudaHostRegister} function can be used to pre-populate the page table on the CPU side. However, we measured the cost of this call to be in the range of an additional 300 ms in the Rodinia application srad. Similar results can also be achieved by adding an artificial pre-initialization loop into the CPU code, which eliminates the overhead of the CUDA API call.

\subsection{System Page Size}
\label{sec:pagesize}

\begin{figure}[bt]
    \centering
    \includegraphics[width=0.9\linewidth]{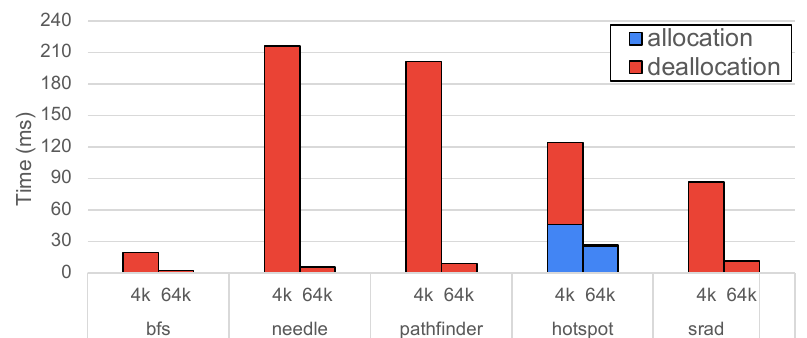}
    \caption{The allocation and de-allocation time in six applications in the system memory version in 64KB and 4KB pages.}
    \label{fig:pagesize-alloc}
\end{figure}
\begin{figure}[bt]
    \centering
    \includegraphics[width=0.9\linewidth]{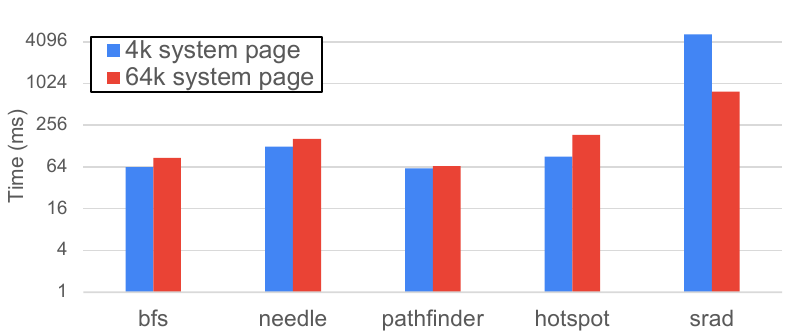}
    \caption{The computation time in six applications in the system memory version in 64~KB and 4~KB pages.}
    \label{fig:pagesize-comp}
\end{figure}

The system page size has impact on both the system allocated memory and CUDA managed memory. All pages in a system allocation use the system page size, while only pages resident on CPU memory in the managed memory uses the system page size. The system page sizes mainly impacts the page initialization overhead that often occurs in application initialization phase, and migration performance between CPU and GPU memory that often occurs in the computation phase. Therefore, we breakdown each application into these two main phases to study the impact of system page sizes.

We run each application in the system memory version by configuring the system pages in 4~KB and 64~KB, respectively. Figure~\ref{fig:pagesize-alloc} compares the allocation and de-allocation time in each Rodina application. A noticeable difference between 4~KB and 64~KB pages lies in the de-allocation time, which is significantly higher in 4~KB system pages, for all applications. Four out of five applications have nearly negligible allocation time. As expected, both allocation and de-allocation time reduce significantly in 64~KB system pages compared to 4~KB system pages ($4.6\times$-$38\times$ with an average of $15.9\times$), as more pages need to be used for the same allocation.

Interestingly, Rodinia applications, with the exception of SRAD, exhibit lower compute time for 4~KB pages compared to 64~KB pages ($1.1\times$-$2.1\times$). Figure~\ref{fig:pagesize-comp} compares the computation time of these applications in the two page sizes. One possible reason for the lower performance in 64~KB pages pages is the granularity of migrated pages may cause amplification, resulting in unused data being migrated. This performance loss could also partially be attributed to the automatic migrations that might incur temporary latency increase when the computation accesses on pages that are being migrated, reducing performance. In Rodinia applications, this is particularly noticeable as applications have a short computation time, where migrated data may not be sufficiently reused. For SRAD, the trend appears to be different, as this particular algorithm perform several iterations on the same data. As such, it can benefit from automatic data migration. Users can tune the threshold for migration (see Section~\ref{sec:bg-access-counters}) to delay page migrations. Detailed results on automatic page migration are presented for this application in Section~\ref{sec:page-migration}.

\begin{figure}[bt]
    \centering
    \includegraphics[width=\linewidth]{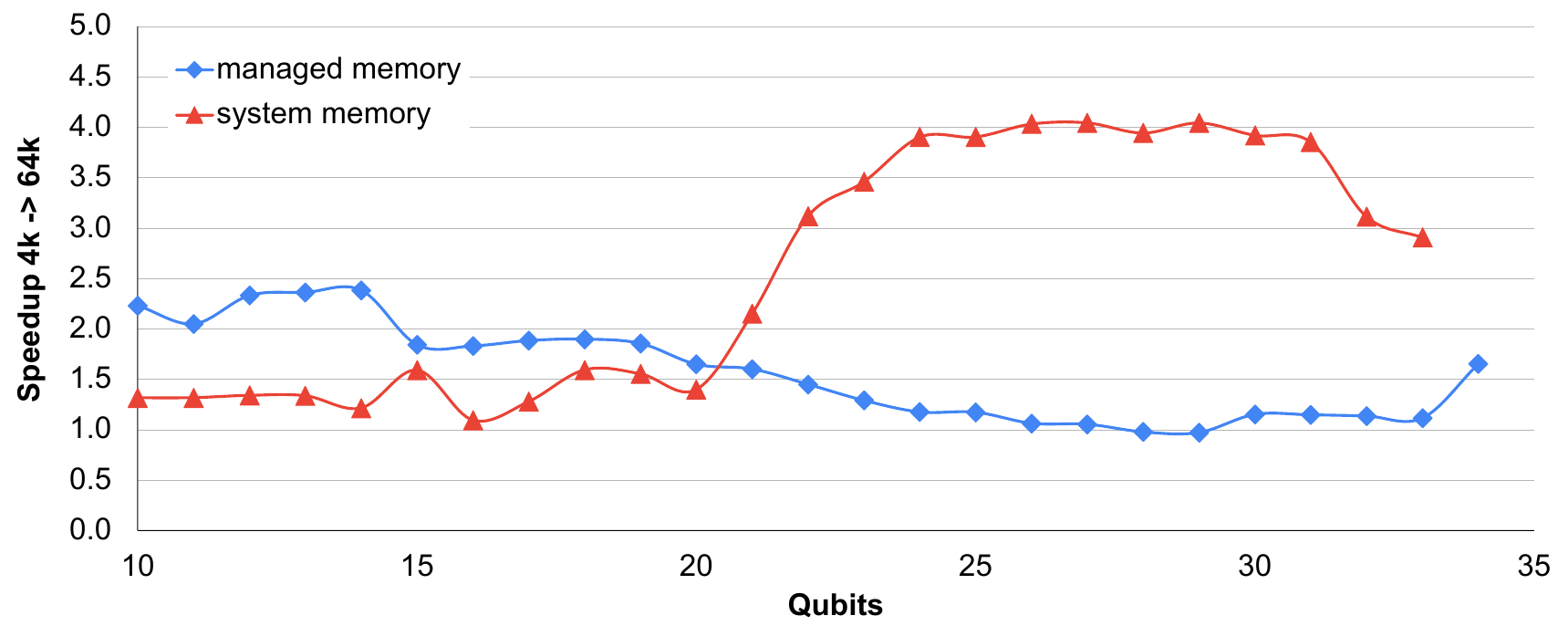}
    \caption{The speedup of Quantum Volume simulations at 64~KB system pages relative to 4~KB system pages in the system memory and the managed memory versions.}
    \label{fig:qiskit_pagesize_speedup}
\end{figure}
\begin{figure}[bt]
    \centering
    \includegraphics[width=0.65\linewidth]{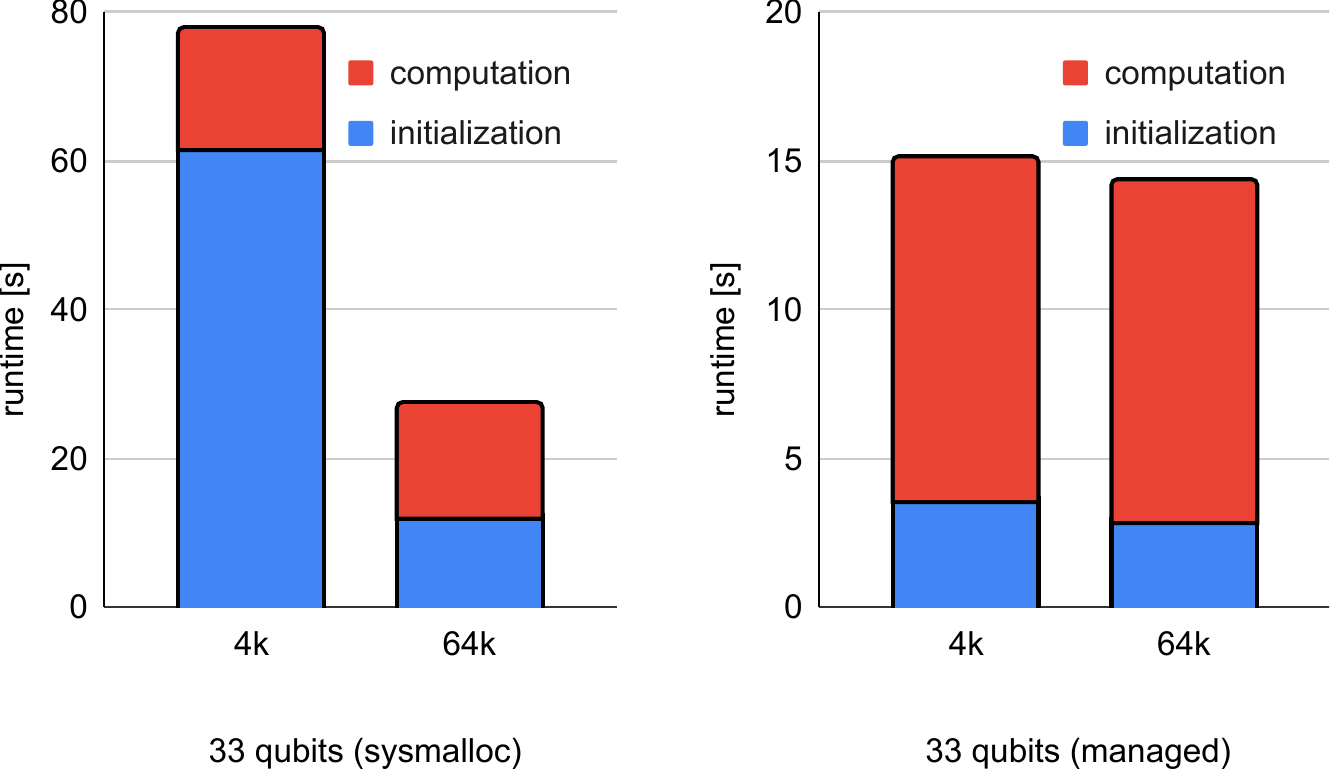}
    \caption{Time breakdown in the initialization and computation phase in a 33-qubit Quantum Volume simulation in the system memory (left) and managed memory (right) versions.}
    \label{fig:qiskit_pagesize}
    \vspace{-10pt}
\end{figure}

Figure~\ref{fig:qiskit_pagesize_speedup} compares the performance of Qiskit Quantum Volume simulations at an increased number of qubits using either 64~KB or 4~KB system pages. For both managed memory and system memory, the system page size has a noticeable impact on performance, bringing up to $2.5\times$ and $4\times$ speedup, respectively. However, with an increasing problem size, the speedup in the managed memory version is decreasing while the speedup in the system memory version is increasing. Starting from 25 qubits, the managed memory version has similar performance in the two page sizes but almost $4\times$ speedup is observed in the system memory version.

In particular, Figure~\ref{fig:qiskit_pagesize} compares the execution time for the 33-qubit case, using 4~KB or 64~KB pages, for both memory allocation methods. In CUDA managed memory, when using 64~KB pages, the execution time is 10\% lower than with 4~KB pages. This limited impact of the system page size is expected, as Qiskit has GPU-side data initialization, and CUDA managed memory uses the GPU page table for GPU-resident data, with a constant 2~MB page size, independent of system page size.

In system memory, the impact of changing the page size to 64~KB is significant, where the overall runtime for 33-qubit is reduced by a factor of $2.9\times$. While the computation time remains stable between page sizes, the initialization time is drastically reduced with 64~KB pages, with a $5\times$ improvement. This difference highlights the cost of GPU-side page initialization, as described in Section~\ref{sec:gpu-first-touch}, where memory pages are first-touched on the GPU-side, and page table initialization is performed on the CPU-side, representing a notable bottleneck in the application.

\section{Page Migration}
\label{sec:page-migration}
\begin{figure}[bt]
    \centering
    \includegraphics[width=\linewidth]{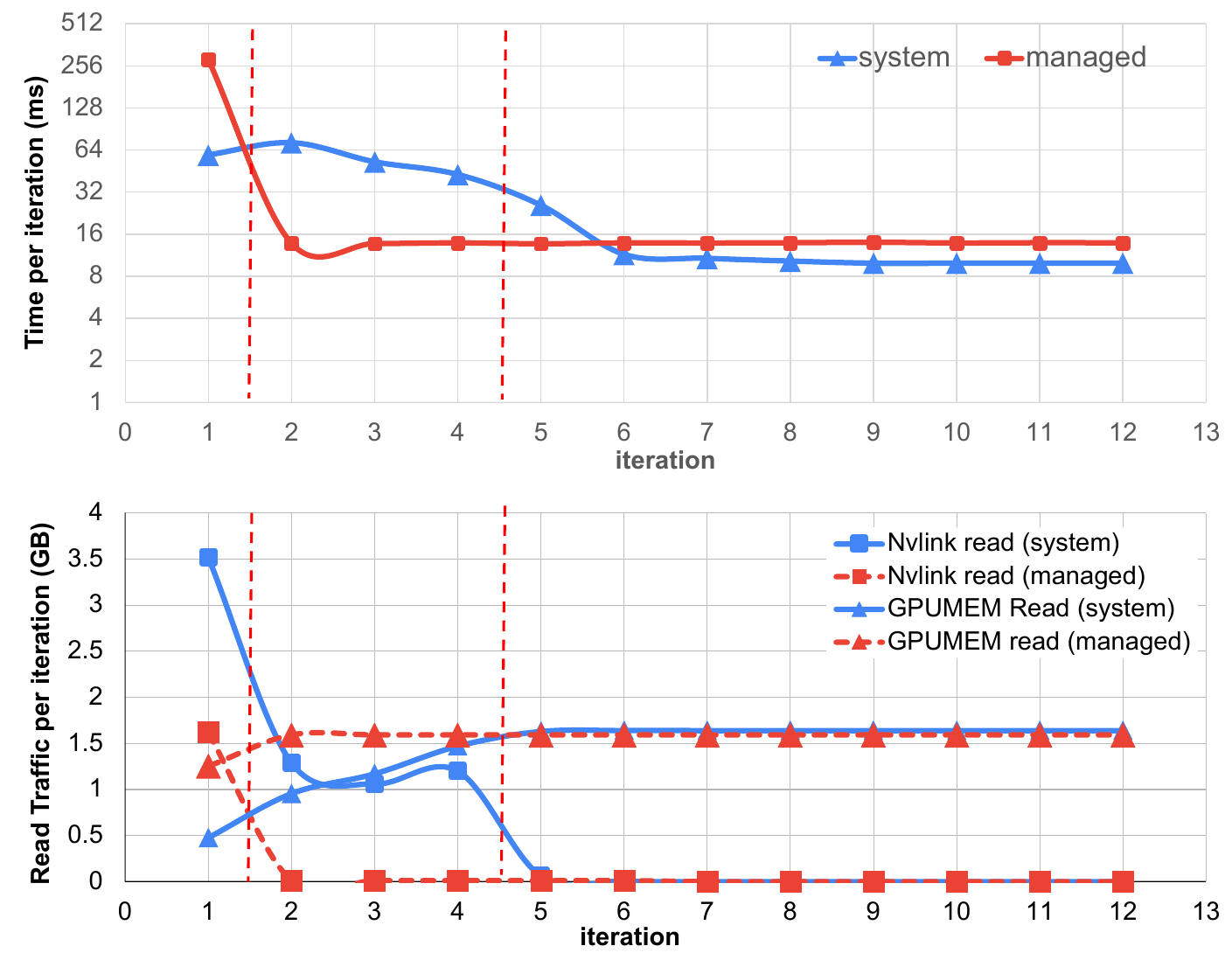}
    \caption{The execution time per iteration  (top) and the memory traffic (bottom) throughout the computational phase of 12 iterations of the SRAD application.}
    \label{fig:srad-migrate}
\end{figure}

In this section, we evaluate the impact of page migrations on Grace Hopper on real applications. In particular, we compare the new automatic access-counter-based strategy in system-allocated memory on Grace Hopper with the on-demand page migration strategy in CUDA managed memory. Experiments in this this section all use 64~KB system pages.%

An application needs to have access patterns that can clearly expose hot pages to exploit the access-counter-based strategy in system-allocated memory. We examined all the test applications and choose SRAD as this application uses an iterative algorithm in its computation phase. Therefore, with a sufficient number of iterations, the access-counter-based page migration should migrate pages repeatedly accessed during computation iterations into GPU memory. Meanwhile, the on-demand page migration in the managed system version should migrate all accessed pages on their first access.

Figure~\ref{fig:srad-migrate} presents the measured execution time for each iteration in SRAD, for the two memory management versions. For the managed memory version, due to page migration in the first iteration, the execution time of this iteration is significantly higher than the other iterations.

In the system memory version, from a performance standpoint, the computational phase consists of three sub-phases, as separated by dashed line on Figure~\ref{fig:srad-migrate}.

The first phase corresponds to the first iteration, with high execution time, primarily caused by the overhead of GPU first-touch on system-allocated data, as memory pages must be initialized on the CPU-side. The second phase (iteration $2$-$4$), exhibits a decreasing iteration time but still slower than that of the managed memory version. In the final phase (iteration $5$ and above), the iteration time stabilizes and outperform the managed memory version.

We further measure the memory traffic in each computation iteration in SRAD and correlate it with the runtime in Figure~\ref{fig:srad-migrate}. For both system-allocated and managed versions, we report the memory read from GPU memory, and remote memory reads over NVLink-C2C. In the managed memory version, all reads are performed from GPU memory, even for the first iteration, where pages are being migrated, and exhibit non-zero reads over NVLink-C2C. This is because in managed memory, pages are first migrated, and then read from local GPU memory. In the system memory version, we observe that memory reads over NVLink-C2C decreases as reads from GPU memory increases gradually in iteration $1$-$4$. This observation confirms that the access-pattern-based automatic migrations are being triggered in this stage, which hinders performance in this period. After the entire working set has been migrated to GPU memory, that is, for iterations $5$-$12$, memory reads over NVLink-C2C remain nearly zero while reads from GPU memory stabilize at 1.5~GB per iteration. Consequently, the performance in iterations $5$-$12$ improves to outperform that of the managed memory version.

For system memory, in SRAD, no memory migration from GPU memory to CPU memory is observed. The main reason is that although some GPU-resident data is read from CPU in computation phase, those reads are not significant enough compared to GPU reads to trigger automatic migration from GPU to CPU. This behavior is also expected as it is not be desirable to migrate pages with low CPU-initiated accesses, where GPU-residency is preferable, while on-demand page migration in the managed memory may cause page thrashing in such scenarios.

\section{Memory Oversubscription}
\label{sec:oversub}
In this section, we study the performance and efficiency of the two unified memory solutions, in memory oversubscription situations. In an oversubscription scenario, the working set of applications exceeds the available GPU memory. Thus, data accessed might not be in GPU memory. Two approaches might address this issue on the Grace Hopper system. First, pages can be evicted from GPU memory, and the required pages can be migrated into GPU. This is the expected behavior for CUDA managed memory. In addition, as Grace Hopper supports direct memory access over NVLink-C2C, data in CPU memory can be remotely accessed without migration.

\begin{figure}[bt]
    \centering
    \includegraphics[width=\linewidth]{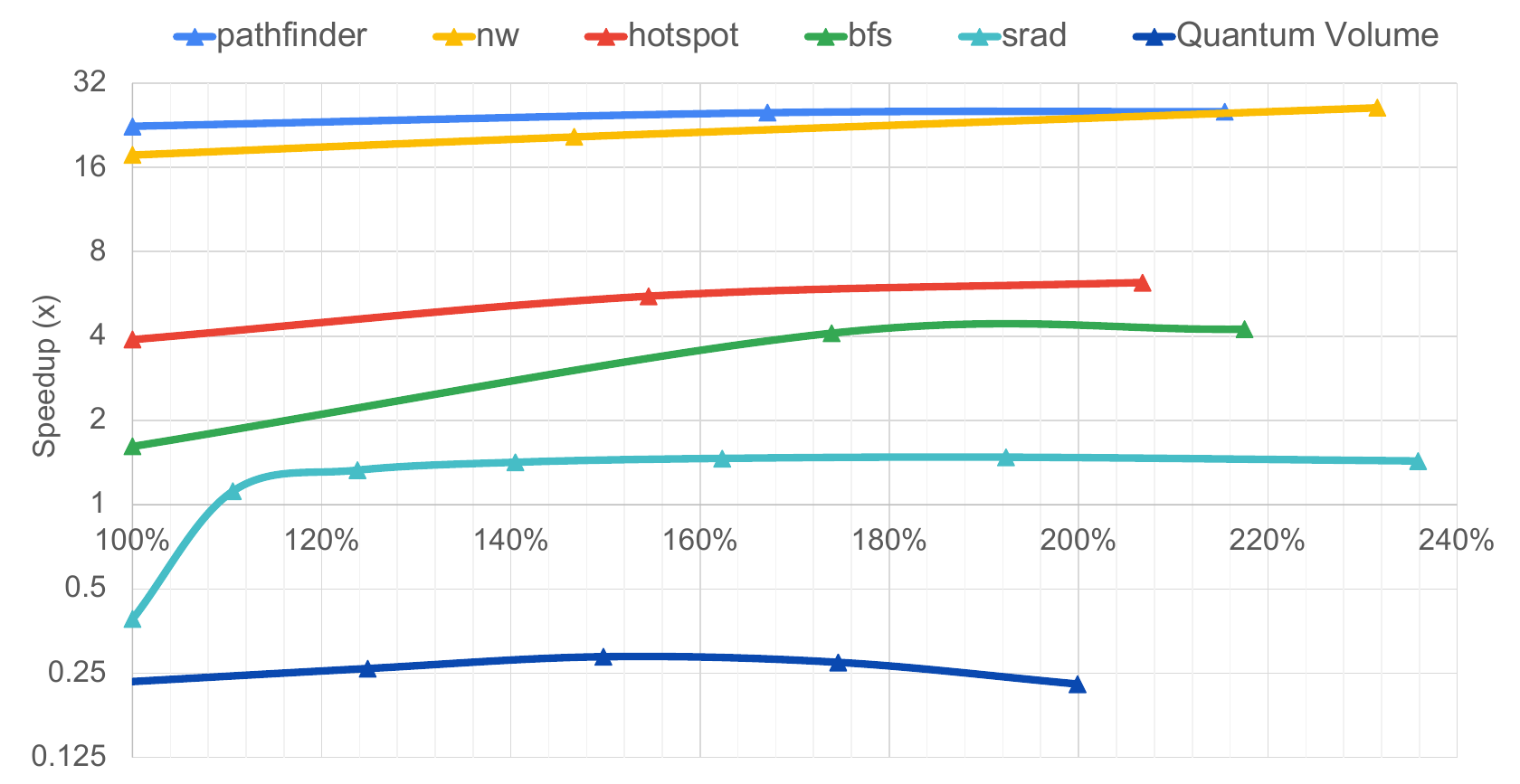}
    \caption{The relative speedup of using system memory compared to managed memory in six applications at increased memory oversubscription.}
    \label{fig:oversub-all}
\end{figure}
\begin{figure}[bt]
    \centering
    \includegraphics[width=\linewidth]{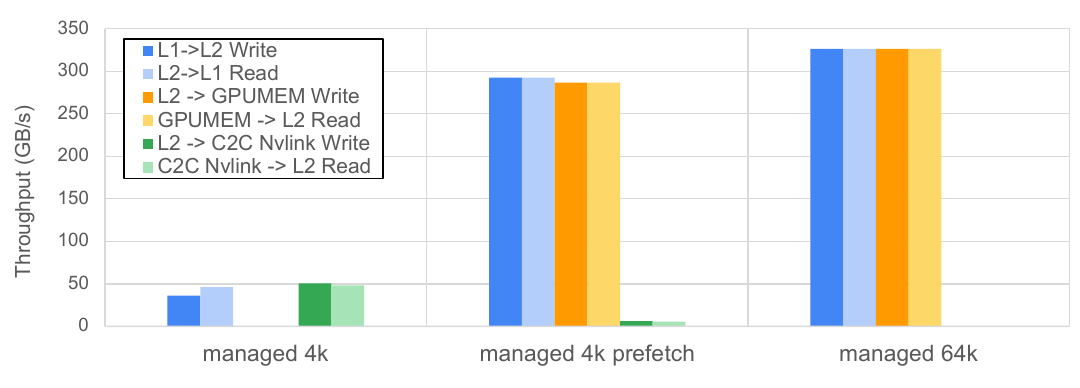}
    \caption{The measured memory throughput in three tiers of the memory hierarchy in the 34 qubit Qiskit quantum volume simulation (130\% GPU memory oversubscription).}
    \label{fig:qiskit_traffic}
\end{figure}
\begin{figure}[bt]
    \centering
    \includegraphics[width=0.75\linewidth]{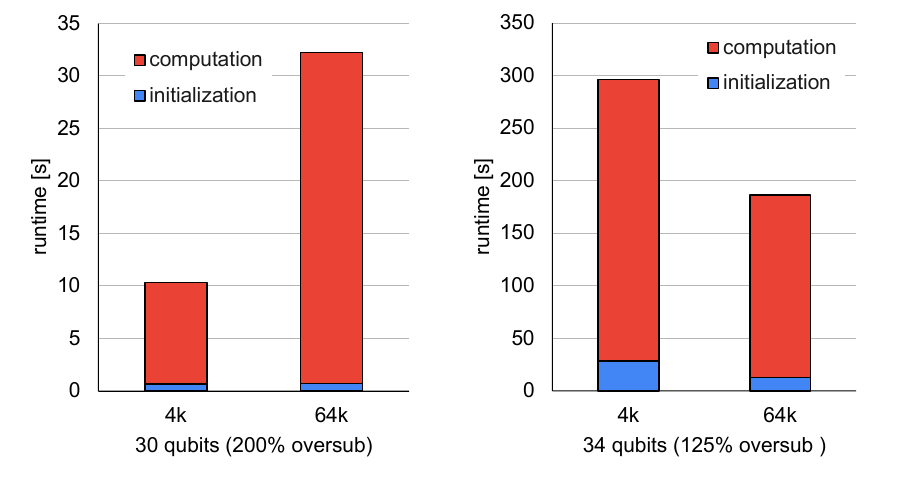}
    \caption{The time breakdown in initialization and computation of a 30-qubit quantum volume simulation (left) and a 34-qubit quantum volume simulation (right).}
    \label{fig:qiskit_oversub}
\end{figure}

We first evaluate the relative speed of the system memory version compared to the managed memory version at increased memory oversubscription. Figure~\ref{fig:oversub-all} presents the results of all six applications. All applications are run with 4~KB system pages. The system memory version of Rodinia applications, BFS, hotspot, needle, pathfinder, are less affected by oversubscription than the managed versions, as indicated by the increased speedup at increased oversubscription. This trend is because that the system-memory version always places data on CPU memory, and performs accesses over NVLink-C2C link. However, in the managed memory version, data is being migrated to the GPU, and evicted when the GPU memory has been exhausted. This eviction and migration process significantly impacts the performance.

Only SRAD exhibits a high impact on its runtime when oversubscribing memory. We further analyze each iteration at different oversubscription ratios. We observe that as oversubscription is increased, the gap between the system memory version and the managed memory version increases. Profiling results suggest that this is induced by large-size page migrations in the managed version, while only small-size remote accesses are performed in the system memory version.

For the 34-qubits quantum volume simulation (about 130\% GPU memory oversubscription), a significant slowdown with managed memory is observed compared to the explicit copy version. Further analysis reveals that no page is migrated and all data is accessed over NVLink-C2C at a low bandwidth. We optimize the managed memory version using CUDA managed memory prefetching to transform the majority of data access to be read locally from GPU memory. As shown in Figure~\ref{fig:qiskit_traffic}, we use memory traffic between L1 and L2 caches as an indication of data rate being fed to the GPU for computation. Clearly, in the managed 4~KB system pages case, the computation is throttled due to the slow NVLink-C2C traffic, which throttles throughput between L1 and L2. After applying the prefetch optimization, the throughput between L1 and L2 mostly is greatly improved, as most throughput comes from GPU memory.

In previous in-memory scenarios, CUDA managed memory in both 4~KB and 64~KB pages exhibits similar execution times. However, in oversubscription scenarios, the system page size shows a high impact on execution time. In the 34-qubit quantum simulation, as shown in Figures~\ref{fig:qiskit_traffic} and~\ref{fig:qiskit_oversub}, switching from 4~KB to 64~KB system pages shortens initialization and accelerates page migration by $58\%$. Interestingly, the 30-qubit simulation shows a different preference on the system page size, nearly $3\times$ slower computation when using 64~KB system pages as shown in Figure~\ref{fig:qiskit_oversub}. This is unexpected, as the page size for GPU-resident memory is 2~MB in managed memory, and is not modified by the system page size. We suggest that this difference is due to some pages being evicted to CPU memory where the system page size is used. In the case of 64~KB pages, when those pages are migrated back to the GPU, the amount of migrated memory at a time is higher than 4~KB, affecting performance.

In case of natural oversubscription in Qiskit (i.e. 34 qubits) with managed memory, we observed the above described eviction happening at the beginning of the simulation for the duration of the initialization phase. Afterwards, no migration (and accordingly no further eviction) is performed throughout the remaining computation phase of the simulation and data is only accessed via the NVLink-C2C link. With manual migrations, i.e., explicit prefetching, the initial eviction is still taking place, but the prefetching causes data to be migrated back into the GPU memory, which results in higher performance, as described earlier.

For system memory, the 34 qubits case, i.e. natural oversubscription, could not be simulated with neither 4~KB nor 64~KB page sizes. Therefore, we use the simulated oversubscription using 30 qubits as a base, which requires approximately 8~GB of memory. Starting with 4~KB pages, when looking at page eviction, we can observe that for the oversubscription scenario with CUDA managed memory, there is a longer page eviction phase observable from GPU memory indicating the initialization phase at the beginning of the run. With system memory, no eviction is happening however.

\section{Related Works}

\textbf{Characterization.}
Early evaluation of the Grace Hopper Superchip from an application perspective has been conducted~\cite{simakov2024first}, along with evaluation of the Grace CPU~\cite{banchelli2024nvidia}. While those work provide performance measurements, utilization and evaluation of the system-allocated memory on Grace Hopper is still widely unexplored. \citet{li2024automatic} propose to leverage automatic page migrations, enabled by the Grace Hopper Superchip, in the context of BLAS computation. Our work focuses on providing an evaluation of the Grace Hopper system-allocated memory using a range of HPC benchmarks, and a state-of-the-art Quantum simulator.

\textbf{CPU-GPU Unified Memory.}
Before Grace Hopper, software-based systems for unified CPU-GPU memory have been extensively studied on Nvidia GPUs.
The overhead of using UVM has been identified to be notably induced by data eviction~\cite{allen2021demystifying}, and page fault handling~\cite{kim2020batch}.
The prefetching mechanism in play in UVM is detailed by \citet{ganguly2019interplay}. Other prefetching strategies have been proposed to improve performance of UVM~\cite{go2023early,long2023deep}. The effectiveness of explicit user-initiated prefetching and software hints to guide data placement has also been evaluated~\cite{chien2019performance}.
\citet{ganguly2020adaptive} proposed a delayed migration strategy, based on memory accesses, to move frequently-accessed pages to the GPU.
\citet{gayatri2019comparing} evaluates the previous generation of NVLink CPU-GPU interconnect by comparing the performance of Unified Memory using Address Translation Service (ATS) with CUDA managed memory.
This work focuses on new hardware-enabled mechanisms available on Grace Hopper, such as the system-level page table and automatic access counter-based page migrations.

\textbf{Heterogeneous Memory System.}
With the emergence of different memory technologies, extensive works have explored various designs of heterogeneous memory systems on HPC platforms~\cite{peng2016exploring,peng2018siena,wahlgren2022evaluating,wahlgren2023quantitative}. Data placement among the multiple memory tiers is crucial to efficiently leverage heterogeneous memory systems. Extensive works have proposed runtime and OS solutions to optimize data movement and memory management in heterogeneous memory systems~\cite{agarwal2017thermostat,yan2019nimble,chen2020atmem,doudali2021cori,maruf2023tpp,hildebrand2024cachedarrays}. These works, in general, aim at placing frequently-accessed pages in higher-performance memory, and rely on either page tables or performance hardware counters to quantify page accesses within acceptable overhead. On Grace Hopper, we focus on its counter-based migration strategy and provide in-depth understanding of this strategy. Other works propose solutions for leveraging heterogeneous memory systems in specific application domains, such as deep learning applications~\cite{ren2021sentinel}, molecular dynamics simulations~\cite{xie2021md}, and plasma physics simulations~\cite{ren2021optimizing}.

\section{Conclusion}
This work provides an in-depth study of integrated CPU-GPU system memory on the Grace Hopper system. We designed and implemented a set of six representative applications, including the state-of-the-art Qiskit quantum computing simulator, using two unified memory management strategies -- system-allocated memory and CUDA-managed memory. Leveraging a memory utilization profiler and hardware counters, we quantified and characterized the impact of the integrated CPU-GPU system page table. Our study focuses on key factors, including first-touch policy, page table entry initialization, page sizes, and page migration, and identifies optimization strategies for different access patterns. Our results show that as a new solution for unified memory, the system-allocated memory can benefit most use cases, with minimal porting efforts. Future works will need a deep understanding of the access counter-based migration on diverse workloads for exploiting this new unified memory system.

\section*{Acknowledgments}
This research is supported by the Swedish Research Council (no. 2022.03062). This work has received funding from the European High Performance Computing Joint Undertaking (JU) and Sweden, Finland, Germany, Greece, France, Slovenia, Spain, and the Czech Republic under grant agreement No.~101093261.

\bibliographystyle{ACM-Reference-Format}
\bibliography{main}


\begin{thebibliography}{34}

%%% the Web address.

\ifx \showCODEN    \undefined \def \showCODEN     #1{\unskip}     \fi
\ifx \showDOI      \undefined \def \showDOI       #1{#1}\fi
\ifx \showISBNx    \undefined \def \showISBNx     #1{\unskip}     \fi
\ifx \showISBNxiii \undefined \def \showISBNxiii  #1{\unskip}     \fi
\ifx \showISSN     \undefined \def \showISSN      #1{\unskip}     \fi
\ifx \showLCCN     \undefined \def \showLCCN      #1{\unskip}     \fi
\ifx \shownote     \undefined \def \shownote      #1{#1}          \fi
\ifx \showarticletitle \undefined \def \showarticletitle #1{#1}   \fi
\ifx \showURL      \undefined \def \showURL       {\relax}        \fi
% invisible to TeX
\providecommand\bibfield[2]{#2}
\providecommand\bibinfo[2]{#2}
\providecommand\natexlab[1]{#1}
\providecommand\showeprint[2][]{arXiv:#2}

\bibitem[Agarwal and Wenisch(2017)]%
        {agarwal2017thermostat}
\bibfield{author}{\bibinfo{person}{Neha Agarwal} {and} \bibinfo{person}{Thomas~F Wenisch}.} \bibinfo{year}{2017}\natexlab{}.
\newblock \showarticletitle{Thermostat: Application-transparent page management for two-tiered main memory}. In \bibinfo{booktitle}{\emph{Proceedings of the Twenty-Second International Conference on Architectural Support for Programming Languages and Operating Systems}}. \bibinfo{pages}{631--644}.
\newblock


\bibitem[Allen and Ge(2021)]%
        {allen2021demystifying}
\bibfield{author}{\bibinfo{person}{Tyler Allen} {and} \bibinfo{person}{Rong Ge}.} \bibinfo{year}{2021}\natexlab{}.
\newblock \showarticletitle{Demystifying gpu uvm cost with deep runtime and workload analysis}. In \bibinfo{booktitle}{\emph{2021 IEEE International Parallel and Distributed Processing Symposium (IPDPS)}}. IEEE, \bibinfo{pages}{141--150}.
\newblock


\bibitem[Banchelli et~al\mbox{.}(2024)]%
        {banchelli2024nvidia}
\bibfield{author}{\bibinfo{person}{Fabio Banchelli}, \bibinfo{person}{Joan Vinyals-Ylla-Catala}, \bibinfo{person}{Josep Pocurull}, \bibinfo{person}{Marc Clasc{\`a}}, \bibinfo{person}{Kilian Peiro}, \bibinfo{person}{Filippo Spiga}, \bibinfo{person}{Marta Garcia-Gasulla}, {and} \bibinfo{person}{Filippo Mantovani}.} \bibinfo{year}{2024}\natexlab{}.
\newblock \showarticletitle{NVIDIA Grace Superchip Early Evaluation for HPC Applications}. In \bibinfo{booktitle}{\emph{Proceedings of the International Conference on High Performance Computing in Asia-Pacific Region Workshops}}. \bibinfo{pages}{45--54}.
\newblock


\bibitem[Che et~al\mbox{.}(2009)]%
        {che2009rodinia}
\bibfield{author}{\bibinfo{person}{Shuai Che}, \bibinfo{person}{Michael Boyer}, \bibinfo{person}{Jiayuan Meng}, \bibinfo{person}{David Tarjan}, \bibinfo{person}{Jeremy~W Sheaffer}, \bibinfo{person}{Sang-Ha Lee}, {and} \bibinfo{person}{Kevin Skadron}.} \bibinfo{year}{2009}\natexlab{}.
\newblock \showarticletitle{Rodinia: A benchmark suite for heterogeneous computing}. In \bibinfo{booktitle}{\emph{2009 IEEE international symposium on workload characterization (IISWC)}}. IEEE, \bibinfo{pages}{44--54}.
\newblock


\bibitem[Chen et~al\mbox{.}(2020)]%
        {chen2020atmem}
\bibfield{author}{\bibinfo{person}{Yu Chen}, \bibinfo{person}{Ivy Peng}, \bibinfo{person}{Zhen Peng}, \bibinfo{person}{Xu Liu}, {and} \bibinfo{person}{Bin Ren}.} \bibinfo{year}{2020}\natexlab{}.
\newblock \showarticletitle{Atmem: Adaptive data placement in graph applications on heterogeneous memories}. In \bibinfo{booktitle}{\emph{Proceedings of the 18th ACM/IEEE International Symposium on Code Generation and Optimization}}. \bibinfo{pages}{293--304}.
\newblock


\bibitem[Chien et~al\mbox{.}(2019)]%
        {chien2019performance}
\bibfield{author}{\bibinfo{person}{Steven Chien}, \bibinfo{person}{Ivy Peng}, {and} \bibinfo{person}{Stefano Markidis}.} \bibinfo{year}{2019}\natexlab{}.
\newblock \showarticletitle{Performance evaluation of advanced features in CUDA unified memory}. In \bibinfo{booktitle}{\emph{2019 IEEE/ACM Workshop on Memory Centric High Performance Computing (MCHPC)}}. IEEE, \bibinfo{pages}{50--57}.
\newblock


\bibitem[Doudali et~al\mbox{.}(2021)]%
        {doudali2021cori}
\bibfield{author}{\bibinfo{person}{Thaleia~Dimitra Doudali}, \bibinfo{person}{Daniel Zahka}, {and} \bibinfo{person}{Ada Gavrilovska}.} \bibinfo{year}{2021}\natexlab{}.
\newblock \showarticletitle{Cori: Dancing to the right beat of periodic data movements over hybrid memory systems}. In \bibinfo{booktitle}{\emph{2021 IEEE International Parallel and Distributed Processing Symposium (IPDPS)}}. IEEE, \bibinfo{pages}{350--359}.
\newblock


\bibitem[Faj et~al\mbox{.}(2023)]%
        {qiskita1002023}
\bibfield{author}{\bibinfo{person}{Jennifer Faj}, \bibinfo{person}{Ivy Peng}, \bibinfo{person}{Jacob Wahlgren}, {and} \bibinfo{person}{Stefano Markidis}.} \bibinfo{year}{2023}\natexlab{}.
\newblock \showarticletitle{Quantum Computer Simulations at Warp Speed: Assessing the Impact of GPU Acceleration: A Case Study with IBM Qiskit Aer, Nvidia Thrust \& cuQuantum}. In \bibinfo{booktitle}{\emph{2023 IEEE 19th International Conference on e-Science (e-Science)}}. \bibinfo{pages}{1--10}.
\newblock
\urldef\tempurl%
\url{https://doi.org/10.1109/e-Science58273.2023.10254803}
\showDOI{\tempurl}


\bibitem[Ganguly et~al\mbox{.}(2019)]%
        {ganguly2019interplay}
\bibfield{author}{\bibinfo{person}{Debashis Ganguly}, \bibinfo{person}{Ziyu Zhang}, \bibinfo{person}{Jun Yang}, {and} \bibinfo{person}{Rami Melhem}.} \bibinfo{year}{2019}\natexlab{}.
\newblock \showarticletitle{Interplay between hardware prefetcher and page eviction policy in cpu-gpu unified virtual memory}. In \bibinfo{booktitle}{\emph{Proceedings of the 46th International Symposium on Computer Architecture}}. \bibinfo{pages}{224--235}.
\newblock


\bibitem[Ganguly et~al\mbox{.}(2020)]%
        {ganguly2020adaptive}
\bibfield{author}{\bibinfo{person}{Debashis Ganguly}, \bibinfo{person}{Ziyu Zhang}, \bibinfo{person}{Jun Yang}, {and} \bibinfo{person}{Rami Melhem}.} \bibinfo{year}{2020}\natexlab{}.
\newblock \showarticletitle{Adaptive page migration for irregular data-intensive applications under gpu memory oversubscription}. In \bibinfo{booktitle}{\emph{2020 IEEE International Parallel and Distributed Processing Symposium (IPDPS)}}. IEEE, \bibinfo{pages}{451--461}.
\newblock


\bibitem[Gayatri et~al\mbox{.}(2019)]%
        {gayatri2019comparing}
\bibfield{author}{\bibinfo{person}{Rahulkumar Gayatri}, \bibinfo{person}{Kevin Gott}, {and} \bibinfo{person}{Jack Deslippe}.} \bibinfo{year}{2019}\natexlab{}.
\newblock \showarticletitle{Comparing managed memory and ats with and without prefetching on nvidia volta gpus}. In \bibinfo{booktitle}{\emph{2019 IEEE/ACM Performance Modeling, Benchmarking and Simulation of High Performance Computer Systems (PMBS)}}. IEEE, \bibinfo{pages}{41--46}.
\newblock


\bibitem[Go et~al\mbox{.}(2023)]%
        {go2023early}
\bibfield{author}{\bibinfo{person}{Seokjin Go}, \bibinfo{person}{Hyunwuk Lee}, \bibinfo{person}{Junsung Kim}, \bibinfo{person}{Jiwon Lee}, \bibinfo{person}{Myung~Kuk Yoon}, {and} \bibinfo{person}{Won~Woo Ro}.} \bibinfo{year}{2023}\natexlab{}.
\newblock \showarticletitle{Early-Adaptor: An Adaptive Framework for Proactive UVM Memory Management}. In \bibinfo{booktitle}{\emph{2023 IEEE International Symposium on Performance Analysis of Systems and Software (ISPASS)}}. IEEE, \bibinfo{pages}{248--258}.
\newblock


\bibitem[Hildebrand et~al\mbox{.}(2024)]%
        {hildebrand2024cachedarrays}
\bibfield{author}{\bibinfo{person}{Mark Hildebrand}, \bibinfo{person}{Jason Lowe-Power}, {and} \bibinfo{person}{Venkatesh Akella}.} \bibinfo{year}{2024}\natexlab{}.
\newblock \showarticletitle{CachedArrays: Optimizing Data Movement for Heterogeneous Memory Systems}. 38th IEEE International Parallel and Distributed Processing Symposium (IPDPS).
\newblock


\bibitem[Kim et~al\mbox{.}(2020)]%
        {kim2020batch}
\bibfield{author}{\bibinfo{person}{Hyojong Kim}, \bibinfo{person}{Jaewoong Sim}, \bibinfo{person}{Prasun Gera}, \bibinfo{person}{Ramyad Hadidi}, {and} \bibinfo{person}{Hyesoon Kim}.} \bibinfo{year}{2020}\natexlab{}.
\newblock \showarticletitle{Batch-aware unified memory management in GPUs for irregular workloads}. In \bibinfo{booktitle}{\emph{Proceedings of the Twenty-Fifth International Conference on Architectural Support for Programming Languages and Operating Systems}}. \bibinfo{pages}{1357--1370}.
\newblock


\bibitem[Li et~al\mbox{.}(2024)]%
        {li2024automatic}
\bibfield{author}{\bibinfo{person}{Junjie Li}, \bibinfo{person}{Yinzhi Wang}, \bibinfo{person}{Xiao Liang}, {and} \bibinfo{person}{Hang Liu}.} \bibinfo{year}{2024}\natexlab{}.
\newblock \showarticletitle{Automatic BLAS Offloading on Unified Memory Architecture: A Study on NVIDIA Grace-Hopper}. In \bibinfo{booktitle}{\emph{Practice and Experience in Advanced Research Computing (PEARC’24)}}.
\newblock


\bibitem[Long et~al\mbox{.}(2023)]%
        {long2023deep}
\bibfield{author}{\bibinfo{person}{Xinjian Long}, \bibinfo{person}{Xiangyang Gong}, \bibinfo{person}{Bo Zhang}, {and} \bibinfo{person}{Huiyang Zhou}.} \bibinfo{year}{2023}\natexlab{}.
\newblock \showarticletitle{Deep learning based data prefetching in CPU-GPU unified virtual memory}.
\newblock \bibinfo{journal}{\emph{J. Parallel and Distrib. Comput.}}  \bibinfo{volume}{174} (\bibinfo{year}{2023}), \bibinfo{pages}{19--31}.
\newblock


\bibitem[Lutz et~al\mbox{.}(2020)]%
        {lutz2020pump}
\bibfield{author}{\bibinfo{person}{Clemens Lutz}, \bibinfo{person}{Sebastian Bre{\ss}}, \bibinfo{person}{Steffen Zeuch}, \bibinfo{person}{Tilmann Rabl}, {and} \bibinfo{person}{Volker Markl}.} \bibinfo{year}{2020}\natexlab{}.
\newblock \showarticletitle{Pump up the volume: Processing large data on gpus with fast interconnects}. In \bibinfo{booktitle}{\emph{Proceedings of the 2020 ACM SIGMOD International Conference on Management of Data}}. \bibinfo{pages}{1633--1649}.
\newblock


\bibitem[Maruf et~al\mbox{.}(2023)]%
        {maruf2023tpp}
\bibfield{author}{\bibinfo{person}{Hasan~Al Maruf}, \bibinfo{person}{Hao Wang}, \bibinfo{person}{Abhishek Dhanotia}, \bibinfo{person}{Johannes Weiner}, \bibinfo{person}{Niket Agarwal}, \bibinfo{person}{Pallab Bhattacharya}, \bibinfo{person}{Chris Petersen}, \bibinfo{person}{Mosharaf Chowdhury}, \bibinfo{person}{Shobhit Kanaujia}, {and} \bibinfo{person}{Prakash Chauhan}.} \bibinfo{year}{2023}\natexlab{}.
\newblock \showarticletitle{Tpp: Transparent page placement for cxl-enabled tiered-memory}. In \bibinfo{booktitle}{\emph{Proceedings of the 28th ACM International Conference on Architectural Support for Programming Languages and Operating Systems, Volume 3}}. \bibinfo{pages}{742--755}.
\newblock


\bibitem[Nvidia(2019)]%
        {unifiedmem}
\bibfield{author}{\bibinfo{person}{Nvidia}.} \bibinfo{year}{2019}\natexlab{}.
\newblock \bibinfo{title}{{Unified Memory}}.
\newblock \bibinfo{howpublished}{\url{https://devblogs.nvidia.com/unified-memory-in-cuda-6/}}.
\newblock


\bibitem[Nvidia(2024a)]%
        {gh_whitepaper}
\bibfield{author}{\bibinfo{person}{Nvidia}.} \bibinfo{year}{2024}\natexlab{a}.
\newblock \bibinfo{booktitle}{\emph{{NVIDIA Grace Hopper Superchip Architecture Whitepaper}}}.
\newblock \bibinfo{type}{Whitepaper}. \bibinfo{institution}{Nvidia}.
\newblock
\newblock
\shownote{Accessed 2024-04-04}.


\bibitem[Nvidia(2024b)]%
        {gh_tuning}
\bibfield{author}{\bibinfo{person}{Nvidia}.} \bibinfo{year}{2024}\natexlab{b}.
\newblock \bibinfo{booktitle}{\emph{{NVIDIA Grace Performance Tuning Guide}}}.
\newblock \bibinfo{type}{{T}echnical {R}eport}. \bibinfo{institution}{Nvidia}.
\newblock
\newblock
\shownote{Accessed 2024-04-01}.


\bibitem[ORNL(2019)]%
        {power9}
\bibfield{author}{\bibinfo{person}{ORNL}.} \bibinfo{year}{2019}\natexlab{}.
\newblock \bibinfo{title}{CUDA Unified Memory}.
\newblock \bibinfo{howpublished}{\url{https://www.olcf.ornl.gov/wp-content/uploads/2019/06/06_Managed_Memory.pdf}}.
\newblock


\bibitem[Pearson et~al\mbox{.}(2019)]%
        {pearson2019evaluating}
\bibfield{author}{\bibinfo{person}{Carl Pearson}, \bibinfo{person}{Abdul Dakkak}, \bibinfo{person}{Sarah Hashash}, \bibinfo{person}{Cheng Li}, \bibinfo{person}{I-Hsin Chung}, \bibinfo{person}{Jinjun Xiong}, {and} \bibinfo{person}{Wen-Mei Hwu}.} \bibinfo{year}{2019}\natexlab{}.
\newblock \showarticletitle{Evaluating characteristics of CUDA communication primitives on high-bandwidth interconnects}. In \bibinfo{booktitle}{\emph{Proceedings of the 2019 ACM/SPEC International Conference on Performance Engineering}}. \bibinfo{pages}{209--218}.
\newblock


\bibitem[Peng et~al\mbox{.}(2016)]%
        {peng2016exploring}
\bibfield{author}{\bibinfo{person}{Ivy~Bo Peng}, \bibinfo{person}{Stefano Markidis}, \bibinfo{person}{Erwin Laure}, \bibinfo{person}{Gokcen Kestor}, {and} \bibinfo{person}{Roberto Gioiosa}.} \bibinfo{year}{2016}\natexlab{}.
\newblock \showarticletitle{Exploring application performance on emerging hybrid-memory supercomputers}. In \bibinfo{booktitle}{\emph{2016 IEEE 18th International Conference on High Performance Computing and Communications; IEEE 14th International Conference on Smart City; IEEE 2nd International Conference on Data Science and Systems}}. IEEE, \bibinfo{pages}{473--480}.
\newblock


\bibitem[Peng and Vetter(2018)]%
        {peng2018siena}
\bibfield{author}{\bibinfo{person}{Ivy~B Peng} {and} \bibinfo{person}{Jeffrey~S Vetter}.} \bibinfo{year}{2018}\natexlab{}.
\newblock \showarticletitle{Siena: Exploring the design space of heterogeneous memory systems}. In \bibinfo{booktitle}{\emph{SC18: International Conference for High Performance Computing, Networking, Storage and Analysis}}. IEEE, \bibinfo{pages}{427--440}.
\newblock


\bibitem[{Qiskit Community}(2017)]%
        {QiskitCommunity2017}
\bibfield{author}{\bibinfo{person}{{Qiskit Community}}.} \bibinfo{year}{2017}\natexlab{}.
\newblock \bibinfo{title}{Qiskit: {{An}} Open-Source Framework for Quantum Computing}.
\newblock
\newblock
\urldef\tempurl%
\url{https://doi.org/10.5281/zenodo.2562110}
\showDOI{\tempurl}


\bibitem[Ren et~al\mbox{.}(2021a)]%
        {ren2021optimizing}
\bibfield{author}{\bibinfo{person}{Jie Ren}, \bibinfo{person}{Jiaolin Luo}, \bibinfo{person}{Ivy Peng}, \bibinfo{person}{Kai Wu}, {and} \bibinfo{person}{Dong Li}.} \bibinfo{year}{2021}\natexlab{a}.
\newblock \showarticletitle{Optimizing large-scale plasma simulations on persistent memory-based heterogeneous memory with effective data placement across memory hierarchy}. In \bibinfo{booktitle}{\emph{Proceedings of the ACM International Conference on Supercomputing}}. \bibinfo{pages}{203--214}.
\newblock


\bibitem[Ren et~al\mbox{.}(2021b)]%
        {ren2021sentinel}
\bibfield{author}{\bibinfo{person}{Jie Ren}, \bibinfo{person}{Jiaolin Luo}, \bibinfo{person}{Kai Wu}, \bibinfo{person}{Minjia Zhang}, \bibinfo{person}{Hyeran Jeon}, {and} \bibinfo{person}{Dong Li}.} \bibinfo{year}{2021}\natexlab{b}.
\newblock \showarticletitle{Sentinel: Efficient tensor migration and allocation on heterogeneous memory systems for deep learning}. In \bibinfo{booktitle}{\emph{2021 IEEE International Symposium on High-Performance Computer Architecture (HPCA)}}. IEEE, \bibinfo{pages}{598--611}.
\newblock


\bibitem[Ren et~al\mbox{.}(2021c)]%
        {atc21:zerooffload}
\bibfield{author}{\bibinfo{person}{Jie Ren}, \bibinfo{person}{Samyam Rajbhandari}, \bibinfo{person}{Reza~Yazdani Aminabadi}, \bibinfo{person}{Olatunji Ruwase}, \bibinfo{person}{Shuangyan Yang}, \bibinfo{person}{Minjia Zhang}, \bibinfo{person}{Dong Li}, {and} \bibinfo{person}{Yuxiong He}.} \bibinfo{year}{2021}\natexlab{c}.
\newblock \showarticletitle{{ZeRO-Offload: Democratizing Billion-Scale Model Training}}. In \bibinfo{booktitle}{\emph{USENIX Annual Technical Conference}}.
\newblock


\bibitem[Simakov et~al\mbox{.}(2024)]%
        {simakov2024first}
\bibfield{author}{\bibinfo{person}{Nikolay~A Simakov}, \bibinfo{person}{Matthew~D Jones}, \bibinfo{person}{Thomas~R Furlani}, \bibinfo{person}{Eva Siegmann}, {and} \bibinfo{person}{Robert~J Harrison}.} \bibinfo{year}{2024}\natexlab{}.
\newblock \showarticletitle{First Impressions of the NVIDIA Grace CPU Superchip and NVIDIA Grace Hopper Superchip for Scientific Workloads}. In \bibinfo{booktitle}{\emph{Proceedings of the International Conference on High Performance Computing in Asia-Pacific Region Workshops}}. \bibinfo{pages}{36--44}.
\newblock


\bibitem[Wahlgren et~al\mbox{.}(2022)]%
        {wahlgren2022evaluating}
\bibfield{author}{\bibinfo{person}{Jacob Wahlgren}, \bibinfo{person}{Maya Gokhale}, {and} \bibinfo{person}{Ivy~B Peng}.} \bibinfo{year}{2022}\natexlab{}.
\newblock \showarticletitle{Evaluating emerging CXL-enabled memory pooling for HPC systems}. In \bibinfo{booktitle}{\emph{2022 IEEE/ACM Workshop on Memory Centric High Performance Computing (MCHPC)}}. IEEE, \bibinfo{pages}{11--20}.
\newblock


\bibitem[Wahlgren et~al\mbox{.}(2023)]%
        {wahlgren2023quantitative}
\bibfield{author}{\bibinfo{person}{Jacob Wahlgren}, \bibinfo{person}{Gabin Schieffer}, \bibinfo{person}{Maya Gokhale}, {and} \bibinfo{person}{Ivy Peng}.} \bibinfo{year}{2023}\natexlab{}.
\newblock \showarticletitle{A quantitative approach for adopting disaggregated memory in HPC systems}. In \bibinfo{booktitle}{\emph{Proceedings of the International Conference for High Performance Computing, Networking, Storage and Analysis}}. \bibinfo{pages}{1--14}.
\newblock


\bibitem[Xie et~al\mbox{.}(2021)]%
        {xie2021md}
\bibfield{author}{\bibinfo{person}{Zhen Xie}, \bibinfo{person}{Wenqian Dong}, \bibinfo{person}{Jie Liu}, \bibinfo{person}{Ivy Peng}, \bibinfo{person}{Yanbao Ma}, {and} \bibinfo{person}{Dong Li}.} \bibinfo{year}{2021}\natexlab{}.
\newblock \showarticletitle{MD-HM: memoization-based molecular dynamics simulations on big memory system}. In \bibinfo{booktitle}{\emph{Proceedings of the ACM International Conference on Supercomputing}}. \bibinfo{pages}{215--226}.
\newblock


\bibitem[Yan et~al\mbox{.}(2019)]%
        {yan2019nimble}
\bibfield{author}{\bibinfo{person}{Zi Yan}, \bibinfo{person}{Daniel Lustig}, \bibinfo{person}{David Nellans}, {and} \bibinfo{person}{Abhishek Bhattacharjee}.} \bibinfo{year}{2019}\natexlab{}.
\newblock \showarticletitle{Nimble page management for tiered memory systems}. In \bibinfo{booktitle}{\emph{Proceedings of the Twenty-Fourth International Conference on Architectural Support for Programming Languages and Operating Systems}}. \bibinfo{pages}{331--345}.
\newblock


\end{thebibliography}

\end{document}